\documentclass[prl,twocolumn,superscriptaddress,preprintnumbers,%
showpacs,amsmath,amssymb]{revtex4}

\usepackage{graphicx}

\newcommand{\N}{\mathcal{N}}

\begin{document}

\title{Finite-temperature spectral function of the vector mesons
  in an AdS/QCD model}
 
\author{Mitsutoshi Fujita}
\affiliation{Department of Physics,
 Kyoto University, Kyoto 606-8502, Japan}

\author{Kenji Fukushima}
\affiliation{Yukawa Institute for Theoretical Physics,
 Kyoto University, Kyoto 606-8502, Japan}

\author{Tatsuhiro Misumi}
\affiliation{Yukawa Institute for Theoretical Physics,
 Kyoto University, Kyoto 606-8502, Japan}

\author{Masaki Murata}
\affiliation{Yukawa Institute for Theoretical Physics,
 Kyoto University, Kyoto 606-8502, Japan}

%\date{\today}

\begin{abstract}
 We use the soft-wall AdS/QCD model to investigate the
 finite-temperature effects on the spectral function in the vector
 channel.  The dissociation of the vector meson tower onto the AdS
 black hole leads to the in-medium mass shift and the width broadening
 in a way similar to the lattice QCD results for the heavy quarkonium
 at finite temperature.  We also show the momentum dependence of the
 spectral function and find it consistent with screening in a hot
 wind.
\end{abstract}

\preprint{KUNS-2195, YITP-09-19}

\maketitle

%%%%%%%%%%   INTRODUCTION   %%%%%%%%%%

\paragraph*{INTRODUCTION} --- The intrinsic properties of hot and
dense matter out of quarks and gluons in quantum chromodynamics (QCD)
have been drawing our attention to a deeper understanding of
strongly-correlated systems.  Such matter has been created at the
Relativistic Heavy-Ion Collider (RHIC) in Brookhaven and widely called
the strongly-correlated quark-gluon plasma or
sQGP~\cite{Gyulassy:2004zy}.  It is generally challenging to attack
non-perturbative physics problems and there is no systematic strategy
established.   Recently powerful techniques have developed based on
the recognition of the gauge/string duality or the bulk/boundary
correspondence~\cite{Maldacena:1997re,Gubser:1998bc,Witten:1998qj},
with which one can treat the strong-coupling regime in the gauge field
theory on the boundary by solving the weak-coupling (or classical in
some limit) string theory in the bulk Anti de~Sitter (AdS) space.

  In sQGP physics the most striking and well-known achievement from
the gauge/string duality is the analytical computation of the shear
viscosity to the entropy density ratio, $\eta/s$, in the $\N=4$
supersymmetric Yang-Mills plasma~\cite{Policastro:2001yc}.  Although
the theory is different from QCD, the obtained result is valuable
under the present circumstance that the non-perturbative QCD
calculation of the shear viscosity is
cumbersome~\cite{Nakamura:2004sy,Meyer:2007ic}.  The smallness of
$\eta/s$ is an important indication of the sQGP because a large
reaction cross-section implies small $\eta$ in gaseous states.
Another suggestive hint to the sQGP has come from the lattice QCD
study on the $J/\psi$ spectral functions at high
temperature~\cite{Umeda:2000ym,Asakawa:2003re,Datta:2003ww}.  There,
it has been found that the mesonic correlation survives even above
$2T_c$ where $T_c$ is the deconfinement critical temperature.

  This melting temperature for the $J/\psi$ meson inferred from the
lattice QCD simulation is not consistent with the potential model
estimate with the Debye screening~\cite{Matsui:1986dk,Mocsy:2007yj},
though the spectral enhancement after melting may explain the lattice
observation~\cite{Mocsy:2007jz}.  We have not yet reached a consensus
on the right interpretation of the $J/\psi$ spectral functions above
$T_c$.  It is most likely that the heavy-quarkonium spectral property
reflects the sQGP nature.  Hence, it would be intriguing to take
advantage of the dual model to see what spectral shape would transpire
non-perturbatively in strongly-correlated systems.

  We remark that the spectral functions and also the meson
dissociation have been nicely discussed by means of the D3/D7
setup~\cite{Mateos:2006nu,Erdmenger:2007cm}.  The correspondence
between the $J/\psi$ spectral function and the holographic outcome is
not transparent, however, because the embedding of the D7 brane in an
AdS-Schwarzschild background leads to a first-order phase transition
at which the dissociation takes place.

  In this Letter we shall make use of a rather phenomenological
approach to QCD inspired by the success of the AdS/CFT correspondence,
which is generally referred to as the AdS/QCD models or the
holographic QCD models~\cite{Erlich:2005qh}.  Particularly, since
$J/\psi$ is a vector meson, we adopt the ``soft-wall''
model~\cite{Karch:2006pv} (see Ref.~\cite{Ghoroku:2005vt} for a
related idea) which is designed to reproduce the Regge
trajectory of the vector mesons (i.e.\ vector meson tower).  Here, we
remark on some related works along the similar line.  In
Ref.~\cite{Ghoroku:2005kg} the meson spectrum has been discussed in
the ``hard-wall'' holographic QCD model at finite temperature.  The
hard-wall model is, however, not quite appropriate to look into the
spectral functions because the infrared (IR) boundary condition has
ambiguity, while the soft-wall model is smooth in the IR side.
Although the meson mass shift has been investigated in
Ref.~\cite{Kim:2007rt} in the soft-wall framework, its relevance to
the lattice QCD results is rather indirect.

  Our aim is to derive the finite-temperature spectral function in
the vector channel from the holographic QCD model in a way that we can
make a direct comparison with the lattice QCD.\ \  As a matter of
fact, as we will discuss, the resulting spectral shape and the
associated mass shift and width broadening are all in qualitative
agreement with the lattice calculations.

%%%%%%%%%%   MODEL SETUP   %%%%%%%%%%

\paragraph*{MODEL SETUP} --- We make a brief review on the formalism
and equations that we are using.  The soft-wall model proposed in
Ref.~\cite{Karch:2006pv} is composed from the gauge fields $A_{L \mu}$
in $\mathrm{SU_L}(N_f)$ and $A_{R \mu}$ in $\mathrm{SU_R}(N_f)$ and
the bi-fundamental matter $X$ whose ultraviolet (UV) behavior controls
the chiral symmetry breaking.

  The essential point in the soft-wall setup is that the
five-dimensional action contains the dilaton field potential so that
the spectrum realizes the linear confinement.  That is, the Lagrangian
density is multiplied by the smooth IR cutoff function,
$e^{-\Phi(z)}$, where $z$ is the fifth coordinate and $\Phi(z)=cz^2$.
Here, $c$ is the only mass dimensional parameter in the model, and we
will later make all other quantities dimensionless in the unit of
$\sqrt{c}$.

  We can decompose the model action into one piece involving the
 vector field, $V=(A_L+A_R)/2$, and the other involving the
 axial-vector field, $A=(A_L-A_R)/2$.  Then, as long as the vector
 $\mathrm{SU_V}(N_f)$ symmetry is left unbroken, the (linearized)
 equation of motion for $V$ decouples from $X$ and $A$.  Thus, we need
 to solve the following equation of motion in the $V_z=0$ gauge,
\begin{equation}
 \begin{split}
 & \partial_z \bigl[ e^{-\Phi(z)}\sqrt{-g}g^{\lambda\lambda} g^{zz}
  \partial_z V_\lambda \bigr] \\
 &\qquad\qquad + e^{-\Phi(z)}\sqrt{-g}g^{\lambda\lambda}
  g^{\mu\nu} \partial_\mu\partial_\nu V_\lambda = 0.
 \end{split}
\label{eq:eom}
\end{equation}
Because the physical degrees of freedom for the massive vector field
have three polarizations, we eliminated one of four components by
imposing $\partial^\mu V_\mu=0$.  Let us now write down the metric
representing the AdS black hole;
\begin{equation}
 g_{\mu\nu}dx^\mu dx^\nu = \frac{L^2}{z^2} \Bigl[
  -f(z)dt^2 + dx^2 + \frac{1}{f(z)}dz^2 \Bigr],
\label{eq:ads_bh}
\end{equation}
where $f(z)=1-z^4/z_h^4$ with the horizon denoted by $z_h$ and $L$
denotes the radius of the AdS space.  The Hawking temperature is read
as $T=1/(\pi z_h)$, which translates into the temperature of a QGP
medium.

  It should be mentioned that the AdS-Schwarzschild metric
(\ref{eq:ads_bh}) undergoes a first-order phase transition to the AdS
metric and the critical temperature is $T_c\simeq 0.492\sqrt{c}$ in
the soft-wall model case~\cite{Andreev:2006eh,Herzog:2006ra}.  The
temperature of our interest is, as we will see later, up to $\sim
0.15\sqrt{c}$. This metric transition is not a serious flaw in our
present setup, however, by the following twofold reasons.  One is that
what we really want to know is simply the effect of the presence of a
medium on the spectrum and not the spectrum in the ground state
determined self-consistently.  For this limited purpose it makes sense
to accommodate (transient) matter by introducing the meta-stable black
hole metric~(\ref{eq:ads_bh}).  The second reasoning is a more
physical one.  If we are interested in $J/\psi$ specifically, it is
possible to construct the $\mathrm{U}(1)$ soft-wall model in the charm
sector alone because the charm current is conserved.  The relevant
scale $\sqrt{c}=\sqrt{c_{J/\psi}}$ should be about four times bigger
than that for $\rho(770)$, which we shall write $\sqrt{c_\rho}$ for
the moment.  Then, $0.15\sqrt{c_{J/\psi}}$ might be a higher
temperature than $T_c$ which is characterized by $\sqrt{c_\rho}$.  We
will go into a more quantitative argument when we show our numerical
results.  To fully justify this hand-waving argument, one should break
$\mathrm{SU_V}(N_f)$ symmetry and carry out the analysis in
Ref.~\cite{Herzog:2006ra} for three light and one heavy flavors.  This
would be an important check to correctly relate $T_c$ to the
heavy-quarkonium dissociation temperature.  Here we postpone resolving
this quantitatively and let us concentrate on demonstrating
qualitatively that the holographic QCD model works well to elucidate
the vector spectral function.

%%%%%%%%%%   SOLUTION   %%%%%%%%%%

\paragraph*{SOLUTION} --- Before solving (\ref{eq:eom}), it would be
helpful to pursue the analogy to the Schr\"{o}dinger equation in
quantum mechanics.  We take a plane-wave form in four dimensional
space-time as usual.  Then, using dimensionless energy $\omega$ and
momentum $q$ in the unit of $\sqrt{c}$ and changing the fifth
coordinate by $\xi=\sqrt{c}z$, we can get rid of $c$ from the equation
of motion.

  Since the presence of matter breaks Lorentz symmetry, the transverse
($\lambda=i=1,2,3$) and the longitudinal ($\lambda=0$) components in
Eq.~(\ref{eq:eom}) lead to distinct differential equations.  The
difference is, nevertheless, not qualitative but quantitative, and so
we shall show the solution of the transverse $V_i$ field only.
\footnote{The longitudinal part has the solution near the horizon as
$V_0(\xi)\to(1-\xi/\xi_h)^{1/2\pm i\sqrt{(\omega\xi_h)^2-4}/4}$, which
is different from Eq.~(\ref{eq:asym}), implying the existence of a
threshold frequency at $\omega\xi_h=2$ or $\omega=2\pi t$.}

%--- figure ---

\begin{figure}
\includegraphics[width=0.92\columnwidth]{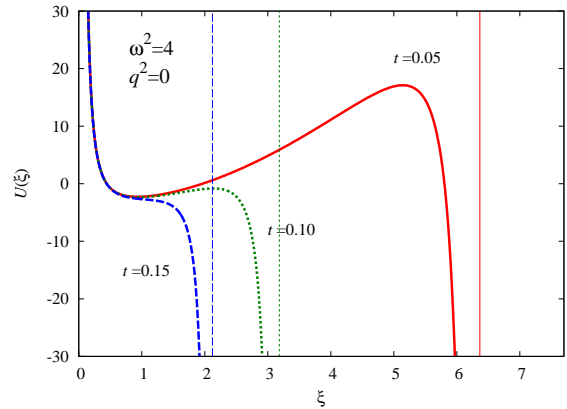}
\caption{Potential $U(\xi)$ for the dimensionless temperatures,
  $t=0.05$, $0.10$, and $0.15$ at $\omega^2=4$ and $q^2=0$.  The thin
  vertical lines represent $\xi_h=1/(\pi t)$, that is the location of
  the black hole horizon.}
\label{fig:poten}
\end{figure}

%--- figure ---

  The change of the field,
$v=(e^{-\Phi}\sqrt{-g}g^{ii}g^{zz})^{1/2}V_i$ (no sum over $i$),
  simplifies the equation of motion in the following form;
$v'' - U(\xi) v=0$ with the potential,
\begin{equation}
 U(\xi) = \xi^2 + \frac{3}{4\xi^2} - \frac{f'}{2f}
  \Bigl( 2\xi + \frac{1}{\xi} \Bigr)
  - \frac{(f')^2}{4f^2} + \frac{f''}{2f}
  - \frac{1}{f} \Bigl( \frac{\omega^2}{f}-q^2 \Bigr).
\label{eq:potential}
\end{equation}
Figure~\ref{fig:poten} shows this potential for various dimensionless
temperatures in the unit of $\sqrt{c}$.  In the $t=0$ case the
downward-convex potential, $\xi^2+3/(4\xi^2)$, yields the discrete
spectrum, $\omega^2=4n$ ($n=1,2,\dots$) for $q^2=0$, only for which
the wave-function is normalizable.  We see that the higher $t$ or
smaller $\xi_h=1/(\pi t)$ makes the potential less convex and
eventually it becomes monotonic for $t\simeq 0.15$.  With a monotonic
potential we cannot expect a remnant of the original spectrum any
more.  In other words we should anticipate dissociation then.

  At finite temperature the potential is no longer rising in the large
$\xi$ side and the normalizability does not quantize the spectrum.
We can easily extract the asymptotic solutions of Eq.~(\ref{eq:eom})
near the horizon as
\begin{equation}
 V_i(\xi) \to \; \phi_{\pm} = (1-\xi/\xi_h)^{\pm i\omega\xi_h/4},
\label{eq:asym}
\end{equation}
as $\xi\to\xi_h$.  Here $\phi_+$ represents the out-coming solution and
$\phi_-$ the in-falling solution into the black hole.  Because the
imaginary part of the \textit{retarded} Minkowskian Green's function
gives the spectral function, the right IR boundary condition should
pick only $\phi_-$ up near the horizon.

%%%%%%%%%%   SPECTRAL FUNCTIONS   %%%%%%%%%%

\paragraph*{SPECTRAL FUNCTIONS} --- Following the procedure elucidated
in Ref.~\cite{Teaney:2006nc} in details, we can numerically get two
independent solutions starting from the UV limit, that is,
$\Phi_1(\xi)$ and $\Phi_2(\xi)$ satisfying $\Phi_1(\xi)\simeq \xi^2$
and $\Phi_2(\xi)\simeq 1$ in the vicinity of $\xi=0$.  Then we can
make an appropriate linear combination,
\begin{equation}
 V_i(\xi) = A(\omega,q)\Phi_2(\xi) + B(\omega,q)\Phi_1(\xi)
  \;\to\; \phi_-(\xi),
\end{equation}
as $\xi\to\xi_h$, with the commonly adopted normalization $A=1$
(i.e.\ $V_i(0)=1$).

%--- figure ---

\begin{figure}
\includegraphics[width=\columnwidth]{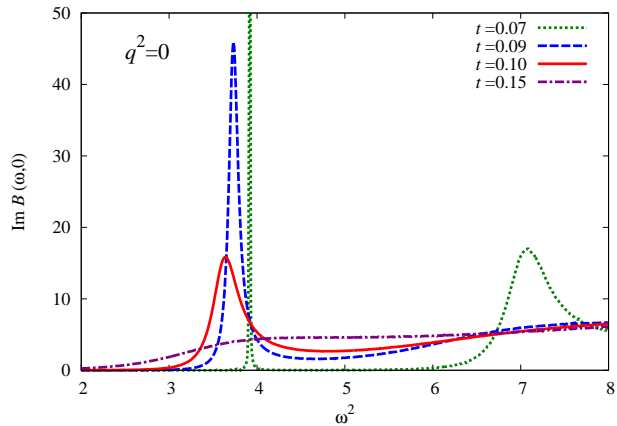}
\caption{Spectral functions $\mathrm{Im}\,B(\omega,0)$ for the
  temperatures, $t=0.07$, $0.09$, $0.10$, and $0.15$.}
\label{fig:spect_t}
\end{figure}

%--- figure ---

  From the prescription for Minkowskian correlators in
Ref.~\cite{Son:2002sd} we can relate the UV property of $V_i(\xi)$ to
the retarded Green's function by
\begin{equation}
 D^{\text{R}}(\omega,q) = -C \lim_{\xi\to 0} \biggl(
  \frac{1}{\xi} V_i^\ast \partial_\xi V_i \biggr) = -2C B(\omega,q),
\end{equation}
where $C$ is a constant given as $N_c/(24\pi^2)$.  The spectral
function is, by its definition,
$\rho(\omega,q)=-(\mathrm{Im}D^{\text{R}}(\omega,q))/\pi
=(2C/\pi)\,\mathrm{Im}B(\omega,q)$.  To be free from the normalization
convention we plot $\mathrm{Im}B(\omega,q)$ as a function of $\omega$
and $q$ for various temperatures.  We also refer to
$\mathrm{Im}B(\omega,q)$ as the spectral function discarding the
overall factor.

  We show our numerical results in Fig.~\ref{fig:spect_t} for $q=0$.
It is obvious at a glance that, at low temperature, sharp peaks stand
in accord with the spectrum $\omega^2=4n$ with $n=1,2,\dots$ known at
$t=0$.  Since the eigenvalues for not $\Phi_1(\xi)$ but $\Phi_2(\xi)$
are relevant to our prescription, the peak positions are slightly
shifted from $\omega^2=4n$, as seen in the data for $t=0.07$ whose
lowest-lying peak is located at $\omega^2=3.92$.  We can identify this
peak as the $J/\psi$ mass squared, from which we get our dimension
unit, $\sqrt{c_{J/\psi}}=1.56\,\text{GeV}$.  We find that the peak
completely vanishes for $t$ as high as $0.15$.  This clearly
corresponds to the temperature where the potential $U(\xi)$ in
Fig.~\ref{fig:poten} loses convexity.  The deconfinement transition
is, on the other hand, $T_c=0.492\sqrt{c_\rho}=0.19\,\text{GeV}$.
This means that in our non-perturbative soft-wall QCD model the
spectral peak melts around
$T\sim 0.15\sqrt{c_{J/\psi}}\simeq 1.2T_c$.  We note, however, that we
should not take this melting temperature in terms of $T_c$ too
seriously because we just borrowed $T_c$ from the estimate in
Ref.~\cite{Herzog:2006ra}.  Our patched-together (decoupled
$\mathrm{U}(1)$ charm sector plus light quarks for $T_c$) argument is
valid only approximately and does not go beyond qualitative estimate.
In principle we can improve the holographic deconfinement scale
provided in the pure gluonic sector by mimicking
QCD~\cite{Gubser:2008ny} and the heavy quarkonium is to be put on top
of it.  Such self-consistent treatment will be reported elsewhere.

  It is notable in Fig.~\ref{fig:spect_t} that the second lowest-lying
state near $\omega^2=8$ melts far earlier than the $\omega^2=4$ state.
This is quite natural because higher excited states are less stable
generally.  In terms of the potential a larger $\omega^2$ causes
stronger absorption into the black hole by the term, $-\omega^2/f^2$
in Eq.~(\ref{eq:potential}), which is negative large near the horizon.
Furthermore, the lowest-lying state moves only slightly to a smaller
mass, while the excited states shift more as seen in the second peak
in the $t=0.07$ curve.

%--- figure ---

\begin{figure}
\includegraphics[width=\columnwidth]{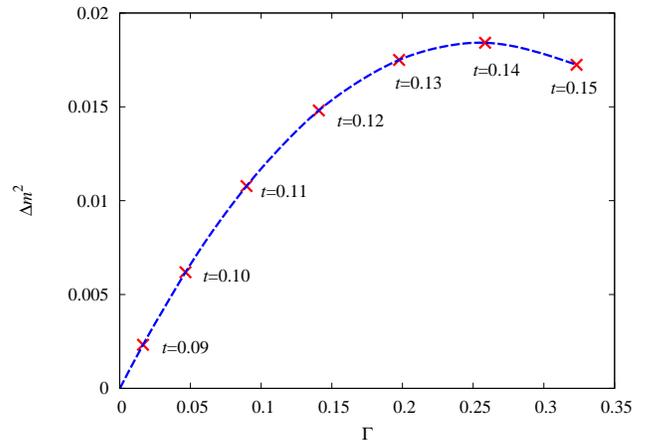}
\caption{Mass shift squared as a function of the width with changing
  temperatures.  The dashed curve smoothly connects the calculated
  points with the interval $\Delta t=0.005$ and the crosses mark
  the points separated by $\Delta t=0.01$.}
\label{fig:mass}
\end{figure}

%--- figure ---

  In Fig.~\ref{fig:mass} we plot the mass shift $\Delta m$ and the
width $\Gamma$ associated with the lowest-lying peak in the spectral
function.  We have found that a functional form,
$a\,\omega^b/[(\omega-\omega_0)^2+\Gamma^2]$, fits the 
spectral shape precisely, from which we extracted the peak position
$\omega_0(t)$ (leading to the mass shift
$\Delta m(t)=\omega_0(0)-\omega_0(t)$) and the width $\Gamma(t)$ as a
function of the temperature.  Interestingly enough, our numerical
results show that $(\Delta m(t))^2\propto\Gamma(t)$ seems to hold as
long as the spectral peak is sharp.  This relation seemingly makes a
contrast to the linear proportionality predicted in the QCD sum
rule~\cite{Morita:2007pt} but could be interpreted as
consistent~\cite{Morita}.  The mass shift is saturated with increasing
$t$, while the width broadening goes on consistently.  The saturation
behavior above $t=0.14$, as depicted in Fig.~\ref{fig:mass}, can
\textit{define} where the vector meson melts quantitatively in terms
of the spectral function.

%--- figure ---

\begin{figure}
\includegraphics[width=1.1\columnwidth]{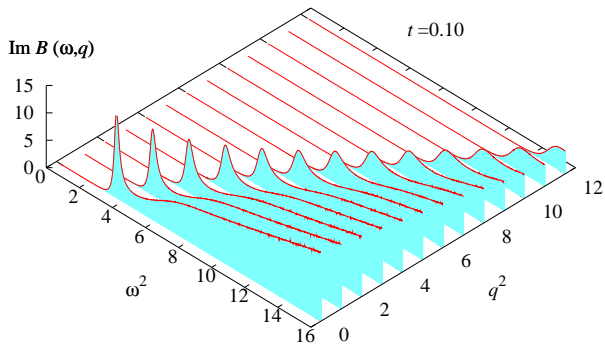}
\vspace{-1cm}
\caption{Spectral functions $\mathrm{Im}B(\omega,q)$ as a function of
  $\omega$ and $q$ for a fixed temperature $t=0.10$.}
\label{fig:spect010}
\end{figure}

%--- figure ---

  Finally we shall discuss the momentum dependence of the spectral
functions.  We plot the numerical results in Fig.~\ref{fig:spect010}
for $q^2$ ranging from $0$ to $12$ with $t=0.10$ fixed.  It is
apparent that the spectral peak is gradually collapsed as $q$
increases.  This is highly non-trivial.  In the perturbative regime,
typically, a larger $q$ makes the spectral peak less sensitive to the
medium effect~\cite{Hidaka:2002xv}.  In strongly-correlated matter, as
discussed in Ref.~\cite{Liu:2006nn} (see also
Ref.~\cite{Myers:2008cj}), $J/\psi$ is more screened and melts at
higher $q$, or in a frame where $J/\psi$ is at rest, it melts under
the hot wind of matter.  Figure~\ref{fig:spect010} clearly illustrates
that the $J/\psi$ suppression by the hot wind is also the case in our
calculation.  It is intriguing that the screening is evident at the
momentum scale $q\sim\sqrt{10}$ which corresponds to about
$5\;\text{GeV}$, which is not far from the transverse momentum of
particles observed at RHIC.

%%%%%%%%%%   SUMMARY   %%%%%%%%%%

\paragraph*{SUMMARY} --- We derived the spectral function of the
vector meson states at finite temperature using the soft-wall AdS/QCD
model.  The qualitative properties of the obtained spectral pattern
are in good agreement with the lattice observation for the heavy
quarkonium.

  We numerically found that the mass shift squared is approximately
proportional to the width broadening.  Another interesting finding is
that the spectral peak diminishes at high momentum.  We can interpret
this as consistent with the $J/\psi$ suppression under the hot wind.

  In our future publication we will report not only the vector channel
addressed here but also the axial vector channel.  It would also be an
important extension to investigate the medium modification with finite
baryon number.  In that case the Chern-Simons coupling plays an
intriguing role in the vector--axial-vector
mixing~\cite{Domokos:2007kt,Harada:2009cn}.

%%%%%%%%%%   ACKNOWLEDGMENTS   %%%%%%%%%%

\begin{acknowledgments}
We are grateful to Toru Kikuchi for discussions and Yoshimasa Hidaka
and Misha Stephanov for comments.  M.~F.\ thanks Kentaroh Yoshida and
Tatsuo Azeyanagi for discussions.  T.~M.\ and M.~M.\ thank Noriaki
Ogawa for technical supports.  K.~F.\ is supported by Japanese MEXT
grant No.\ 20740134 and also supported in part by Yukawa International
Program for Quark Hadron Sciences.
\end{acknowledgments}

\end{document}